\documentclass [12pt]{article}
\pdfoutput=1

\usepackage{amsmath}
\usepackage{amsfonts}
\usepackage{amscd}
\usepackage{amsthm}
\usepackage{setspace}

\usepackage{graphicx}
\usepackage{authblk}
\usepackage{caption}
\usepackage{ytableau}
\usepackage{mathtools}

\def\Z{\mathbb{Z}}

\def\C{\mathbb{C}}
\def\P{\mathbb{P}}

\begin{document}

\begin{titlepage}

\begin{flushright}
KEK-TH-2303
\end{flushright}

\vskip 3.0cm

\begin{center}

{\large New perspectives in the duality of M-theory, heterotic strings, and F-theory}

\vskip 1.2cm

Yusuke Kimura$^1$ 
\vskip 0.6cm
{\it $^1$KEK Theory Center, Institute of Particle and Nuclear Studies, KEK, \\ 1-1 Oho, Tsukuba, Ibaraki 305-0801, Japan}
\vskip 0.4cm
E-mail: kimurayu@post.kek.jp

\vskip 2cm
\abstract{We discuss a new perspective on the dualities among seven-dimensional M-theory on elliptically fibered K3 surfaces, eight-dimensional (8D) heterotic strings on $T^2$, and 8D F-theory on elliptic K3 surfaces. There are several distinct small-fiber F-theory limits of a single M-theory, and we deduce that the distinct F-theory limits of an identical M-theory generally include both an F-theory limit, wherein a discrete gauge symmetry forms, and another F-theory limit, wherein a discrete gauge symmetry does not form. 
\par We also discuss constraints imposed on the degrees of discrete gauge groups and on the continuous gauge groups formed in F-theory on K3 surfaces by applying a formula that is known to hold for genus-one fibrations of K3 surfaces, and by utilizing the existence of the Jacobian fibrations.}  

\end{center}
\end{titlepage}

\tableofcontents
\section{Introduction}
The web of dualities among the versions of superstring theories, M-theory \cite{Witten1995}, and F-theory have been discussed, and numerous efforts have been made to test dualities among the theories. In this note, we discuss a novel aspect of these dualities. In F-theory \cite{Vaf, MV1, MV2}, compactification spaces have genus-one fibrations. However, given a space that admits a genus-one fibration, the choice of the genus-one fibration structure is not unique; for example, elliptic K3 surfaces generally admit several distinct elliptic fibrations. In this study, we interpret this degree of freedom when choosing distinct elliptic fibration structures occurring in the F-theory moduli from the M-theoretic and heterotic perspectives. 
\par In this work, we discuss eight-dimensional (8D) F-theory on elliptic K3 surfaces, dual seven-dimensional (7D) M-theory on K3 surfaces, and dual 8D heterotic strings on 2-torus $T^2$.
\par There are two goals in this study:
\par One goal is to interpret the degree of freedom when choosing several elliptic fibration structures of an elliptic K3 surface when the complex structure of that K3 surface is fixed from M-theoretic and heterotic perspectives.
\par The other goal is to introduce a method to constrain the degree of a discrete gauge group and to determine the potential for continuous gauge groups that can form in F-theory on an elliptic K3 surface, utilizing lattice structures of the K3 surface. This second theme is related to the first theme, as we will explain briefly. 
\par We utilize a lattice-theoretic approach to deal with the two areas. This approach has the following advantage: when the complex structure of a K3 surface is known, it does not necessarily yield the equation of an elliptic fibration of the K3 surface. Furthermore, if one elliptic fibration of an elliptic K3 surface is obtained, it is not simple to deduce the equations for other elliptic fibrations from that equation. It is difficult to obtain the Weierstrass equations of distinct elliptic fibration structures of a K3 surface with a known complex structure owing to these difficulties. There are lattice-theoretic techniques that avoid these difficulties when analyzing elliptic fibration structures of elliptic K3 surfaces without relying on the standard Weierstrass techniques, which enable us to deal with these two issues. 

\vspace{5mm}

\par With respect to the first goal, given an elliptically fibered K3 surface with a fixed complex structure, it is not simple to determine all genus-one fibrations of that K3 surface. Generally, an elliptic K3 surface admits both elliptic fibrations with a global section and genus-one fibrations without a global section. Fortunately, when an elliptic K3 surface has Picard number 12 or larger, a mathematical method is known to classify all elliptic fibrations {\it with a global section}. This method is known as the Kneser--Nishiyama method \cite{Nish, Nish97}. The lattice structures of K3 surfaces contain essential geometric information of the surfaces. The Kneser--Nishiyama method uses a specific lattice, which is referred to as the transcendental lattice of a K3 surface, to classify all types of elliptic fibrations with a global section of the K3 surface. 
\par Two distinct elliptic fibrations of a K3 surface with a fixed complex structure have different singularity types and different Mordell--Weil groups; therefore, different gauge groups form in F-theory on distinct elliptic fibrations of a K3 surface with a fixed complex structure \footnote{This fact was used to explore statistically the four-dimensional (4D) F-theory flux vacua on K3 $\times$ K3 in \cite{BKW}.}. Furthermore, a discrete gauge group forms in F-theory on a genus-one fibration lacking a global section \cite{MTsection} \footnote{For recent progress of F-theory on genus-one fibrations without a section and discrete gauge groups in F-theory can also be found, e.g., in \cite{BM, AGGK, KMOPR, GGK, MPTW, MPTW2, BGKintfiber, CDKPP, LMTW, Kimura2015, Kimura201603, Kimura201607, Kimura201608, Kimura201905, Kimura201907, Kimura201908}.}, but a discrete gauge group does not form in F-theory on an elliptic fibration with a section \footnote{Studies of F-theory models on elliptic fibrations admitting a global section can be found, for example, in \cite{MorrisonPark, MPW, BMPWsection, AL, MM, LSW, CKPT, MPT1610, Kimura1802, MizTani2018, LW1905, Kimura1910, CKS1910, Chabrol2019, Kimura1911, FKKMT1912, Kan2020, Karozas2020, Angelantonj2020, Kuramochi2020, He202009, Clingher2020}.}. 
\par When the complex structure of an elliptic K3 surface is fixed, M-theory on that K3 surface yields a single theory; however, the gauge group formed in F-theory on that K3 surface depends on a fibration structure chosen among the elliptic/genus-one fibrations that the K3 surface admits. These distinct theories, which are obtained as F-theory on distinct fibration structures, correspond to different small-fiber F-theory limits of a single M-theory on the K3 surface. 
\par We find that distinct small-fiber F-theory limits of an identical M-theory on an elliptic K3 surface generally include both a theory wherein a discrete gauge group forms, and a theory wherein a discrete gauge symmetry does not form, as we discuss in section \ref{subsec3.2}.

\par The Kneser--Nishiyama method applies to complex K3 surfaces with Picard numbers greater than or equal to 12 \cite{SchShio}. The K3 surface with Picard number 20 is called an attractive K3 surface. Attractive K3 surfaces have the characteristic property that their complex structures are determined by the transcendental lattices \cite{PS-S1971, SI1977}, and the structures of the attractive K3 surfaces enable us to analyze in detail the physics of F-theory on the surfaces. We also choose one attractive K3 surface as a sample, and explicitly apply the method to F-theory on this surface to demonstrate how the method works. Our proposed method applies to M-theory and F-theory on other attractive K3 surfaces. The method also applies to K3 surfaces with other Picard numbers greater than or equal to 12 after appropriate modifications.
\par Distribution of the gauge groups over 4D F-theory flux vacua over K3 $\times$ K3 \footnote{Studies of M-theory and F-theory on K3 $\times$ K3 can also be found, for example, in \cite{Dasgupta1999, Aspinwall2005, Kimura2015, Kimura201603, Braun202009, Bena202010}.} utilizing the Kneser--Nishiyama method was studied in \cite{BKW}. We utilize the lattice theoretic approach and the Kneser--Nishiyama method to analyze the dualities of M-theory, heterotic strings, and F-theory moduli in the context of the present study.  
\par An application of the Kneser--Nishiyama method reveals that every attractive K3 surface admits two elliptic fibrations with a section, whose singularity types include $E_8^2$ and $D_{16}$, respectively \cite{BKW}. Using an argument similar to that given in \cite{KRES, Kimura201810}, we explain that the fibration with the singularity type that includes $E_8^2$ can be seen as a stable degeneration \cite{FMW, AM} \footnote{Heterotic/F-theory duality \cite{Vaf, MV1, MV2, Sen, FMW} is strictly formulated when the stable degeneration is considered on the F-theory side. Recent progress on the stable degenerations can be found, e.g., in \cite{Donagi2012, AHK, BKW, BKL, MizTan2016, Kimura201810}.}, wherein a K3 surface degenerates into the sum of two extremal rational elliptic surfaces intersecting along an elliptic curve. 
\par The two fibrations whose singularity types include $E_8^2$ and $D_{16}$ are two distinct fibrations of an identical attractive K3 surface. They correspond to $E_8\times E_8$ and $SO(32)$ heterotic strings, respectively, where every attractive K3 surface admits two elliptic fibrations whose singularity types include $E_8^2$, and $D_{16}$ implies that limits of an M-theory on an attractive K3 surface includes 8D $E_8\times E_8$ and $SO(32)$ heterotic theories. 
\par This can be seen as an indication of the well-known property that the moduli spaces of $E_8\times E_8$ and $SO(32)$ heterotic strings are connected upon toroidal compactification \cite{Narain1985, Narain1986}. A related discussion can be found in \cite{Candelas1997}. 

\vspace{5mm}

\par The second theme concerns discrete gauge groups formed in F-theory on K3 surfaces. The Kneser--Nishiyama method only applies to elliptic fibrations with a section of K3 surfaces, and genus-one fibrations lacking a global section of a K3 surface cannot be studied using this method. Thus, the method does not provide a way to analyze the discrete gauge symmetries formed in F-theory on K3 surfaces. 
\par Still, some other lattice theoretic techniques \footnote{Studies of lattice structures of the geometry and applications to string theory can be found, for example, in \cite{KRES, Kimura201810, Font2020, Cota202012}.} can be used to investigate the geometries of genus-one fibrations of K3 surfaces without a section, and discrete gauge symmetries that form in F-theory on K3 surfaces can be studied. We utilize the determinants of the transcendental lattices of K3 surfaces and the existence of the Jacobians \footnote{Construction of the Jacobian of an elliptic curve is discussed in \cite{Cas}.} to impose constraints on the discrete and continuous gauge groups formed in F-theory. In some cases, these conditions impose strong constraints on the degrees of possible discrete gauge groups that can form in F-theory compactifications on K3 surfaces. Utilizing both the discriminant of a K3 surface and the existence of the Jacobian can be a novel approach to studying a discrete gauge group formed in F-theory. 
\par In this work, we focus on F-theory on attractive K3 surfaces to apply this approach. A characteristic property of attractive K3 surfaces whereby their complex structures are labeled by triplets of integers simplifies some of the arguments, and this property makes the analysis more transparent. After some small adjustments, the approach also applies to F-theory on K3 surfaces of other Picard numbers.
\par A K3 genus-one fibration always has the Jacobian fibration. An original genus-one fibration and the Jacobian fibration have the identical singularity types, and their discriminant loci are identical, but their complex structures are generally different. Considering the relation of their discriminants impose constraints on the gauge group formed in F-theory on the genus-one fibration. 
\par We impose constraints on the degree of the potential discrete gauge symmetry that can form in F-theory by utilizing a lemma in \cite{Keum} \footnote{This lemma follows from result in \cite{TataMuk}. See also \cite{Bel} for a related discussion.} on a relation that holds between the Picard lattices of a genus-one fibered K3 surface and the Jacobian surface in section \ref{sec4}.

\par Local model constructions \cite{DWmodel, BHV, BHV2, DWGUT} have been emphasized in recent F-theory model buildings. However, because studying the geometry from the global perspective is necessary to discuss the issues pertaining to gravity and the early universe, we analyze the geometry globally in this work. It is widely known in mathematics that the lattice theoretic approach is useful when analyzing the geometric structures of elliptic K3 surfaces. In this work, we utilize a lattice theoretic approach to analyze the structures of elliptic K3 surfaces. We employ dualities to apply the results to M-theory and heterotic strings.

\vspace{5mm}

\par In section \ref{sec2}, we briefly review attractive K3 surfaces, the Kneser--Nishiyama method and extremal rational elliptic surfaces. We apply these notions to study physics of M-theory, heterotic strings and F-theory in sections \ref{sec3} and \ref{sec4}. 
\par In section \ref{sec3}, we discuss from M-theoretic and heterotic perspectives via the dualities the fact one observes in the F-theory moduli on K3 surface that, even when the complex structure of K3 surface is fixed, there remains freedom in choosing genus-one/elliptic fibration structures of the surface. A relation to the stable degeneration limit is also discussed. Summary of the results is stated in section \ref{subsec3.1}. 
\par In section \ref{subsec3.4}, we choose as a sample a point in the complex structure moduli of 8D F-theory that corresponds to an attractive K3 surface, and we demonstrate how different F-theory limits with distinct gauge groups can be obtained from an identical M-theory. We do not exhaustively compute all the elliptic fibrations with a section of the sample K3 surface. Exhaustive computations of the elliptic fibrations with a section of some attractive K3 surfaces can be found, e.g. in \cite{Nish, BLe2011} \footnote{\cite{BKW2013} also computed elliptic fibrations with a section of K3 surfaces via the Kneser--Nishiyama method. Another method to compute elliptic fibrations of a K3 surface can be found in \cite{Besier2019}.}, and a method similar to those in the references yields all the elliptic fibrations with a section of the sample K3 surface that we discuss in this work. We only discuss a few representative elliptic fibrations with a section that allow for consistency checks.
\par In section \ref{sec4}, we discuss that lattice structures of K3 surfaces and the existence condition of the Jacobian fibration impose constraints on discrete gauge symmetries and continuous gauge groups formed in F-theory. We find that for some cases these conditions uniquely determine the degree of the potential discrete gauge symmetry that can form in F-theory. 
\par We state our concluding remarks and we also mention problems left unsolved in section \ref{sec5}.

\section{Reviews of attractive K3 surfaces and lattice-theoretic techniques}
\label{sec2}
We briefly review attractive K3 surfaces, extremal fibrations, and the Kneser--Nishiyama method. These play a role in the analysis of the dualities of M-theory, heterotic strings, and F-theory in section \ref{sec3}, and we discuss constraints on discrete and continuous gauge groups in F-theory in Section \ref{sec4}. 

\subsection{Attractive K3 surfaces}
\label{subsec2.1}
Attractive K3 surfaces are K3 surfaces with Picard number twenty \footnote{We follow the convention of the term utilized in \cite{Moore04} to call such a K3 surface an attractive K3 in this note.}. Generally, K3 surfaces of Picard number $\rho$ has complex structure moduli of dimension $20-\rho$; therefore, the attractive K3 surfaces yield a set of discrete points in the complex structure moduli of whole algebraic complex K3 surfaces. 
\par The complex structures of the attractive K3 surfaces are known to be labelled by triplets of integers modulo a group action of $GL_2(\Z)$ \cite{PS-S1971, SI1977}. The orthogonal complement of N\'eron-Severi lattice embedded inside the K3 lattice, $H^2({\rm K3},\Z)$, is referred to as the transcendental lattice. For an attractive K3 surface, the transcendental lattice is a positive-definite even lattice, whose intersection matrix is given by size-two integral matrix of the following form:
\begin{equation}
\label{intersection matrix attractive in 2.1}
\begin{pmatrix}
2a & b \\
b & 2c \\
\end{pmatrix},
\end{equation}
where integers $a,b,c$ satisfy the relation $a\ge c\ge b\ge 0$. The complex structure of an attractive K3 surface is uniquely specified by the transcendental lattice \cite{PS-S1971, SI1977}. When the transcendental lattice $T_S$ of an attractive K3 $S$ has intersection matrix (\ref{intersection matrix attractive in 2.1}), we denote this attractive K3 by $S_{[2a \hspace{1mm} b \hspace{1mm} 2c]}$ in this note. 
\par Generally an elliptic K3 surface is known to possess multiple distinct elliptic fibrations. Fixing the complex structure of a K3 surface does not specify the theory when F-theory is compactified on a K3 surface; there still remains the freedom in choosing several distinct elliptic fibrations, and distinct fibrations yield different theories. Distinct elliptic fibrations of an elliptic K3 surface (with a fixed complex structure) have different singularity types and different Mordell--Weil groups. The gauge groups formed in F-theory on distinct fibrations are different. Additionally, when an elliptic K3 surface admits a genus-one fibration lacking a global section, one can also consider F-theory on this genus-one fibration. 
\par When a complex K3 surface $S$ has Picard number twenty and the Mordell--Weil group of an elliptic fibration with a section, $e: S\rightarrow \P^1$, is finite, then the fibration $e$ is referred to as an extremal fibration. The complex structures, singularity types and the Mordell--Weil groups of the extremal K3 surfaces were completely classified by the authors in \cite{SZ}. We analyze an extremal K3 surface as an illustrative example in section \ref{subsec3.4}. 
\par We utilize in section \ref{sec3} the following equality that holds for elliptic K3 surfaces $S$ with a section by the Shioda--Tate formula \cite{Shiodamodular, Tate1, Tate2}:
\begin{equation}
\rho(S) = 2+ {\rm rk} \, M + {\rm rk} \, {\rm MW}(S),
\end{equation}
wherein $\rho(S)$ denotes the Picard number of the K3 surface $S$, $M$ represents the root lattice generated by the components of the reducible singular fibers not meeting the zero-section, and rk MW($S$) denotes the rank of the Mordell--Weil group of the surface $S$.
\par The singular fiber types of an elliptic surface were classified, and j-invariants and the monodromies of the singular fibers were also deduced, by Kodaira \cite{Kod1, Kod2}, and the authors in \cite{Ner, Tate} discussed techniques to yield the singular fiber types of an elliptic surface. The singular fiber types of an elliptic surface, j-invariants and the monodromies of the singular fibers are listed in Table \ref{table of fiber types in 2.1}.

\begingroup

\begin{table}[htb]
  \begin{tabular}{|c|c|r|c|c|} \hline
Type of singular fiber & j-invariant & monodromy  & order of monodromy  & singularity type \\ \hline
$I_0^*$ & regular & $-\begin{pmatrix}
1 & 0 \\
0 & 1 \\
\end{pmatrix}$ & 2 & $D_4$\\ \hline
$I_n$ & $\infty$ & $\begin{pmatrix}
1 & n \\
0 & 1 \\
\end{pmatrix}$ & infinite & $A_{n-1}$\\
$I_m^*$ & $\infty$ & $-\begin{pmatrix}
1 & m \\
0 & 1 \\
\end{pmatrix}$ & infinite & $D_{4+m}$\\ \hline
$II^*$ & 0 & $\begin{pmatrix}
0 & -1 \\
1 & 1 \\
\end{pmatrix}$ & 6 & $E_8$\\ 
$II$ & 0 & $\begin{pmatrix}
1 & 1 \\
-1 & 0 \\
\end{pmatrix}$ & 6 & none.\\ \hline
$III^*$ & 1728 & $\begin{pmatrix}
0 & -1 \\
1 & 0 \\
\end{pmatrix}$ & 4 & $E_7$\\ 
$III$ & 1728 & $\begin{pmatrix}
0 & 1 \\
-1 & 0 \\
\end{pmatrix}$ & 4 & $A_1$\\ \hline
$IV^*$ & 0 & $\begin{pmatrix}
-1 & -1 \\
1 & 0 \\
\end{pmatrix}$ & 3 & $E_6$\\ 
$IV$ & 0 & $\begin{pmatrix}
0 & 1 \\
-1 & -1 \\
\end{pmatrix}$ & 3 & $A_2$\\ \hline
\end{tabular}
\caption{\label{table of fiber types in 2.1}Types of the singular fibers, their j-invariants, monodromies, and the corresponding singularity types \cite{Kod1, Kod2}. ``Regular'' for type $I_0^*$ fiber indicates that the j-invariant of $I_0^*$-type fiber can take any finite value in $\C$.}
\end{table}
\endgroup

\par We study some extremal fibrations of an attractive K3 surface in section \ref{sec3} that are obtained as a deformation of the stable degeneration limit. We will find in section \ref{subsec3.4} that one of such extremal fibrations of K3 surface can be viewed as a deformation of the sum of two extremal rational elliptic surfaces. Here, extremal rational elliptic surfaces are rational elliptic surfaces with a section whose Mordell--Weil groups are finite. The extremal rational elliptic surfaces were classified in \cite{MP1986}. For the reader's convenience, the classification result \cite{MP1986} of the extremal rational elliptic surfaces is shown in Table \ref{tab of extremal rational elliptic in 2.1}. The complex structure of an extremal rational elliptic surface is uniquely determined by the types of the singular fibers except the surface $X_{[0^*, \hspace{1mm} 0^*]}(j)$ with two type $I_0^*$ fibers, as discussed in \cite{MP1986}. (The complex structure of the surface $X_{[0^*,  \hspace{1mm} 0^*]}(j)$ depends on the j-invariant, $j$, of the fiber of the elliptic fibration.) Owing to this fact, the extremal rational elliptic surface with a type $II^*$ fiber and a type $II$ fiber, for example, is denoted as $X_{[II^*, \hspace{1mm} II]}$. $X_{[II^*, \hspace{1mm} 1, 1]}$ is used to represent the extremal rational elliptic surface with a type $II^*$ fiber and two type $I_1$ fibers. We simply use $n$ to represent a type $I_n$ fiber, and $m^*$ is employed to represent a type $I^*_{m}$ fiber in the notation. The same notational convention was employed in \cite{KRES}. 

\begingroup

\begin{table}[htb]
\centering
  \begin{tabular}{|c|c|c|} \hline
$
\begin{array}{c}
\mbox{extremal  rational}\\
\mbox{elliptic surface} 
\end{array}
$ & singular fiber type & singularity type \\ \hline
$X_{[II^*, \hspace{1mm} II]}$ & $II$, $II^*$ & $E_8$  \\ \hline
$X_{[III^*, \hspace{1mm} III]}$ & $III$, $III^*$ & $E_7 A_1$  \\ \hline
$X_{[IV^*, \hspace{1mm} IV]}$ & $IV$, $IV^*$ & $E_6 A_2$ \\ \hline
$X_{[0^*, \hspace{1mm} 0^*]}(j)$ & $I_0^*$, $I^*_0$ & $D_4^2$ \\ \hline
$X_{[II^*, \hspace{1mm} 1, 1]}$ & $II^*$, $I_1$, $I_1$ & $E_8$ \\ \hline
$X_{[III^*, \hspace{1mm} 2, 1]}$ & $III^*$, $I_2$, $I_1$ & $E_7 A_1$ \\ \hline
$X_{[IV^*, \hspace{1mm} 3, 1]}$ & $IV^*$, $I_3$, $I_1$ & $E_6 A_2$ \\ \hline
$X_{[4^*, \hspace{1mm} 1, 1]}$ & $I_4^*$, $I_1$, $I_1$ & $D_8$ \\ \hline
$X_{[2^*, \hspace{1mm} 2, 2]}$ & $I_2^*$, $I_2$, $I_2$ & $D_6 A_1^2$ \\ \hline
$X_{[1^*, \hspace{1mm} 4, 1]}$ & $I^*_1$, $I_4$, $I_1$ & $D_5 A_3$ \\ \hline
$X_{[9, \, 1, \, 1, \, 1]}$ & $I_9$, $I_1$, $I_1$, $I_1$ & $A_8$ \\ \hline
$X_{[8, \, 2, \, 1, \, 1]}$ & $I_8$, $I_2$, $I_1$, $I_1$ & $A_7 A_1$ \\ \hline 
$X_{[6, \, 3, \, 2, \, 1]}$ & $I_6$, $I_3$, $I_2$, $I_1$ & $A_5 A_2 A_1$ \\ \hline
$X_{[5, \, 5, \, 1, \, 1]}$ & $I_5$, $I_5$, $I_1$, $I_1$ & $A_4^2$  \\ \hline
$X_{[4, \, 4, \, 2, \, 2]}$ & $I_4$, $I_4$, $I_2$, $I_2$ & $A_3^2 A_1^2$ \\ \hline
$X_{[3, \, 3, \,  3, \, 3]}$ & $I_3$, $I_3$, $I_3$, $I_3$ & $A_2^4$ \\ \hline
\end{tabular}
\caption{\label{tab of extremal rational elliptic in 2.1}Singular fiber types and singularity types of the extremal rational elliptic surfaces \cite{MP1986} are shown.}
\end{table}  
\endgroup

\subsection{Kneser--Nishiyama method}
\label{subsec2.2}
In this note, we utilize lattice-theoretic approaches, particularly a method called the Kneser--Nishiyama method \cite{Nish, Nish97}, to analyze the structures of elliptic K3 surfaces. The method is used to study the dualities of M-theory, heterotic strings and F-theory. We briefly review the Kneser--Nishiyama method here. When a K3 surface admits an elliptic fibration with a section, the method provides a way to determine all the elliptic fibrations with a global section. The method does not yield a genus-one fibration lacking a global section when a K3 surface admits such genus-one fibration. The Kneser--Nishiyama method determines the singular fibers and Mordell--Weil groups of elliptic fibrations with a section.
\par Because we utilize some lattice-theoretic terms in the discussions of sections \ref{sec3} and \ref{sec4}, we review some lattice-theoretic language here. By lattice, we indicate a free $\Z$-module of a finite rank equipped with a symmetric integral non-degenerate bilinear form. When the square of every element $x$ in $L$, $x^2=x\cdot x$, is even, the lattice $L$ is referred to as an even lattice . When $M$ is a sublattice of the lattice $L$, the orthogonal complement of $M$ inside $L$, $M^{\perp}$, is the sublattice of $L$ consisting of the elements in $L$ that are orthogonal to every element in $M$; namely, the orthogonal complement $M^{\perp}$ consists of the elements in $L$, $y$, that have $x\cdot y=0$ for every element $x$ in $M$. When $M$ embeds inside the lattice $L$, $M\subset L$, the embedding is said to be primitive when the quotient lattice, $L/M$, is a free $\Z$-module. For a bases $\{e_i\}$ of the lattice $L$, the determinant of an intersection matrix $(e_{i}\cdot e_{j})_{i j}$ is referred to as the discriminant of the lattice $L$, disc $L$. When the lattice $L$ has the discriminant $\pm 1$, it is said to be unimodular. The unimodular lattices include the hyperbolic plane $H$ and $E_8$ lattice.
\par The Kneser--Nishiyama method applies to complex K3 surfaces of Picard numbers greater than or equal to 12 \cite{SchShio}. The Kneser--Nishiyama method works as follows: When a K3 surface $S$ has the transcendental lattice $T_S$, we consider the primitive embedding of the lattice $T_S [-1]$ into $E_8\oplus H^l$, where $H$ denotes the hyperbolic plane, whose intersection matrix is given as $\begin{pmatrix}
0 & 1 \\
1 & 0 \\
\end{pmatrix}$, and when the K3 surface $S$ has Picard number $\rho$, $l=20-\rho$. When the K3 surface $S$ is attractive, then $l=0$. Then, we take the orthogonal complement of $T_S [-1]$ inside $E_8\oplus H^l$; to line up with the notation used in \cite{Nish}, we denote the orthogonal complement as $T_0$. Once $T_0$ is obtained, one can deduce elliptic fibrations of the K3 surface $S$.
\par When $T_0$ is computed, the Kneser--Nishiyama method deduces the elliptic fibrations of the K3 $S$ by primitively embedding $T_0$ into {\it the Niemeier lattices}. The negative-definite unimodular even lattices of rank 24 are referred to as the Niemeier lattices, and they are classified into 24 isometric types \cite{Niem}. Each type is uniquely determined by the root lattice. The root lattices of the 24 types of Niemeier lattices are shown in Table \ref{table Niemeier root types in 2.2}. 

\begingroup
\renewcommand{\arraystretch}{1.1}
\begin{table}[htb]
\begin{center}
  \begin{tabular}{|c|c|} \hline
$E_8^3$ & $D_5^2A_7^2$ \\
$E_8 D_{16}$ & $D_4^6$ \\
$E_7^2D_{10}$ & $D_4A_5^4$ \\
$E_7A_{17}$ & $A_{24}$ \\
$E_6^4$ & $A_{12}^2$ \\
$E_6D_7A_{11}$ & $A_8^3$ \\
$D_{24}$ & $A_6^4$ \\
$D_{12}^2$ & $A_4^6$ \\
$D_9A_{15}$ & $A_3^8$ \\  
$D_8^3$ & $A_2^{12}$ \\
$D_6^4$ & $A_1^{24}$ \\
$D_6A_9^2$ & 0 \\ \hline
\end{tabular}
\caption{24 root types of the Niemeier lattices as classified in \cite{Niem}. The lattice $E_8^3$ is itself a Niemeier lattice.}
\label{table Niemeier root types in 2.2}
\end{center}
\end{table}  
\endgroup

The inequivalent ways that one can primitively embed $T_0$ into the Niemeier lattices precisely correspond to the types of the elliptic fibrations with a section that the K3 surface $S$ admits, as described in \cite{Nish}. When $T_0$ is primitively embedded into a Niemeier lattice, the root lattice of the orthogonal complement yields the singularity type of an elliptic fibration of K3 $S$. One can also compute the Mordell--Weil group from the orthogonal complement. This method is exhaustive, and one can deduce all elliptic fibrations of K3 $S$ using this method \cite{Nish} \footnote{There always exists an elliptic fibration with a section for each primitive embedding of $T_0$ into a Niemeier lattice when a complex K3 surface has a Picard number greater than or equal to 12 \cite{SchShio}.}. There are 24 types of Niemeier lattices, and for each type, there are several ways to embed $T_0$ into a Niemeier lattice. Thus, an elliptic K3 surface generally has several elliptic fibrations with a section. 
\par When the Picard number of a K3 surface is small, the rank of $T_0$ becomes large and applying the Kneser--Nishiyama method becomes difficult. This method is particularly useful when the Picard number of a K3 surface is large. When a K3 surface is attractive, the rank of $T_0$ is six. 
\par The Kneser--Nishiyama method was applied to several K3 surfaces of high Picard numbers, and $T_0$ and elliptic fibrations are deduced for these K3 surfaces in \cite{Nish}. These K3 surfaces include $S_{[2 \hspace{1mm} 1 \hspace{1mm} 2]}$, namely the attractive K3 surface whose transcendental lattice has the intersection matrix $\begin{pmatrix}
2 & 1 \\
1 & 2 \\
\end{pmatrix}$. This attractive K3 surface has the smallest discriminant, which is three \cite{KimuraMizoguchi}.  
\par $T_0$ of the attractive K3 surface $S_{[2 \hspace{1mm} 1 \hspace{1mm} 2]}$ is isometric to $E_6$ \cite{Nish}. Because $E_6$ is a root lattice, when this $T_0$ primitively embeds into a Niemeier lattice, it embeds inside the root lattice. There are only six Niemeier lattices into which $E_6$ primitively embeds. Owing to this, there are six elliptic fibrations for the attractive K3 $S_{[2 \hspace{1mm} 1 \hspace{1mm} 2]}$, and their singularity types are: $E_8^2 A_2$, $E_7D_{10}$, $E_6^3$, $D_{16}A_2$, $D_7A_{11}$, $A_{17}$. Two fibrations have Mordell--Weil rank one, and the other four fibrations have Mordell--Weil rank zero. The six elliptic fibration types of the surface $S_{[2 \hspace{1mm} 1 \hspace{1mm} 2]}$ and their Mordell--Weil groups are shown in Table \ref{tab fibrations of surface discriminant 3 in 2.2}. 

\begingroup
\renewcommand{\arraystretch}{1.1}
\begin{table}[htb]
\centering
  \begin{tabular}{|c|c|c|} \hline
Fibration type & singularity type & MW group \\ \hline
no.1 & $E^2_8 \, A_2$ & 0 \\ 
no.2 &  $D_{16} \, A_2$ & $\Z_2$ \\
no.3 & $E_7 \, D_{10}$ & $\Z\oplus \Z_2$ \\
no.4 & $A_{17}$ & $\Z\oplus \Z_3$ \\ 
no.5 & $E^3_6$ & $\Z_3$ \\
no.6 & $D_7 \, A_{11}$ & $\Z_4$ \\ \hline
\end{tabular}
\caption{\label{tab fibrations of surface discriminant 3 in 2.2}Elliptic fibration types of the attractive K3 surface $S_{[2 \hspace{1mm} 1 \hspace{1mm} 2]}$, their singularity types and their Mordell--Weil groups \cite{Nish}.}
\end{table}
\endgroup

Because the attractive K3 $S_{[2 \hspace{1mm} 1 \hspace{1mm} 2]}$ has the smallest discriminant, it does not have a genus-one fibration without a section \cite{Keum}. Physics of F-theory on some of these fibrations was studied in \cite{BKW, KimuraMizoguchi}. 
\par Attractive K3 $S_{[2 \hspace{1mm} 1 \hspace{1mm} 2]}$ also appears as the Jacobian fibration of a genus-one fibration in section \ref{sec4}.
\par For situations where $T_0$ is reducible or $T_0$ is not a root lattice, the computations of the Kneser--Nishiyama method become more complicated. When $T_0$ is reducible, the number of ways in which it embeds into the Niemeier lattices increases, yielding an increased number of types of elliptic fibrations.

\section{Dualities of 7D M-theory on K3, 8D F-theory on K3, and 8D heterotic strings on $T^2$}
\label{sec3}

\subsection{Summary of results}
\label{subsec3.1}
We discuss a new perspective on the dualities of 7D M-theory on elliptic K3 surface, and 8D heterotic strings on $T^2$ and 8D F-theory on elliptic K3 surfaces. Generally, an elliptic K3 surface with a fixed complex structure admits several distinct elliptic fibrations with a section and genus-one fibrations without a global section. They have distinct singularity types; therefore, F-theory on these distinct fibrations yields different theories with different gauge groups. F-theory on an elliptic K3 surface depends not only on its complex structure, but the theory also depends on the fibration structure chosen among the freedom of multiple fibrations that the K3 surface possesses.
\par On the heterotic side, these distinct theories over a fixed complex structure on the F-theory moduli correspond to different theories, and on the heterotic side, these can be seen as mainly coming from the freedom of the choices of the vector bundles over $T^2$. 
\par However, on the M-theory side, the theory only depends on the complex structure of an elliptic K3 surface; different theories on the F-theory side coming from the choices of fibration structures for an elliptic K3 surface with a fixed complex structure corresponds to a single theory. From the M-theoretic perspective, these distinct theories arising on the F-theory side can be viewed as the different small-fiber F-theory limits of a single theory on the M-theory side. Given an M-theory on an elliptic K3 surface, there are several ways to take small-fiber limit of F-theory, and the resulting F-theory limit is not unique; there are several different F-theory limits. This freedom when choosing different small-fiber F-theory limits precisely corresponds to the distinct fibration structures of an elliptic K3 surface with a fixed complex structure. 
\par A discrete gauge group arises on a genus-one fibration lacking a global section \cite{MTsection}, while a discrete gauge group does not arise on an elliptic fibration with a global section. As stated previously, an elliptic K3 surface generally admits both an elliptic fibration with a section and a genus-one fibration without a section. Therefore, starting from an M-theory on K3, a discrete gauge group arises in F-theory limit on a genus-one fibration lacking a global section of that K3 surface, while a discrete gauge group does not form in the F-theory limit on an elliptic fibration with a global section of that K3 surface. Because the different fibration structures of an elliptic K3 surface correspond to the ways of taking the F-theory limits, we conclude that starting from an M-theory on an elliptic K3 surface, the formation of a discrete gauge group in an F-theory limit depends on the way that one takes an F-theory limit of the M-theory on the K3 surface.
\par We also find that on the heterotic side, the different F-theory limits originating from a single M-theory include both $E_8\times E_8$ and $SO(32)$ heterotic theories. This is a manifestation of the well-known fact that the moduli spaces of these two versions of the heterotic theories are connected upon toroidal compactification \cite{Narain1985, Narain1986}. The relationship with the stable degeneration limit is also discussed. 
\par We discuss these in sections \ref{subsec3.2} - \ref{subsec3.4} by analyzing the fibrations structures of elliptic K3 surfaces using a mathematical method called the Kneser--Nishiyama method. This technique is reviewed in Section \ref{subsec2.2}. 

\subsection{M-theoretic perspective}
\label{subsec3.2}
We study F-theory on distinct elliptic/genus-one fibrations of an elliptic K3 surface when its complex structure is fixed from the M-theoretic perspective. M-theory on an elliptic K3 surface with a fixed complex structure is a unique theory, but F-theory on the distinct fibrations of the K3 surface yields different theories with different gauge groups. These different theories on the F-theory side can be viewed as different small-fiber F-theory limits of a single M-theory. Distinct elliptic fibrations with a section of an elliptic K3 surface with a fixed complex structure have different singularity types and different Mordell--Weil groups, depending on the fibration structures. The non-Abelian gauge groups and the numbers of U(1) formed in F-theory differ for distinct elliptic fibrations with a section of an elliptic K3 surface when its complex structure is fixed \cite{BKW}. In other words, taking different F-theory limits of an identical M-theory leads to theories with distinct gauge groups on the F-theory side. 
\par Furthermore, an elliptic K3 surface generally admits both an elliptic fibration with a section and a genus-one fibration without a global section. A discrete gauge symmetry does not form in F-theory on the former fibration, whereas a discrete gauge symmetry forms in F-theory on the latter fibration. They also lead to distinct F-theory limits. Starting from an identical M-theory, a discrete gauge symmetry arises in some F-theory limits, but a discrete gauge symmetry does not arise in the other F-theory limits. 
\par The Kneser--Nishiyama method \cite{Nish, Nish97} applied to a K3 surface yields the elliptic fibrations with a section exhaustively, as we reviewed in section \ref{subsec2.2}. The Kneser--Nishiyama method applies to any complex K3 surface with a Picard number greater than or equal to 12 \cite{SchShio}. 
\par Given a K3 surface with a Picard number greater than or equal to 12, one can compute $T_0$ from the transcendental lattice, and the orthogonal complements of the primitive embeddings of $T_0$ into the Niemeier lattices exhaustively yield the elliptic fibrations with a section of that K3 surface; this is known as the Kneser--Nishiyama method. Because there are 24 Niemeier lattices, and there are generally several different ways to embed $T_0$ into a Niemeier lattice, this method yields several distinct elliptic fibration types for a K3 surface with a fixed complex structure. Because the method exhaustively yields all elliptic fibrations with a section of a K3 surface with Picard number greater than or equal to 12, this implies that a general K3 surface of Picard number greater than or equal to 12 has multiple distinct elliptic fibrations with a section. 
\par Distinct elliptic fibrations with a section of a K3 surface that has a fixed complex structure have different singularity types and different Mordell--Weil groups; F-theory compactifications on such distinct elliptic fibrations yield different non-Abelian gauge groups and different numbers of U(1) factors \cite{BKW}. They correspond to different small-fiber F-theory limits of a single M-theory on the K3 surface.
\par The Kneser--Nishiyama method provides a method to determine the gauge groups formed in F-theory limits on elliptic fibrations with a section of a K3 surface (when it has a Picard number greater than or equal to 12). 
\par We provide an illustration of these in section \ref{subsec3.4}. 

\par Furthermore, as we stated previously, a K3 surface generally admits a genus-one fibration lacking a global section as well as an elliptic fibration with a section. This implies that a discrete gauge symmetry forms in a certain F-theory limit of an M-theory on a K3 surface, but a discrete symmetry does not form in other F-theory limits of the same M theory. Whether a discrete symmetry forms depends on the F-theory limit that one takes. We demonstrate this point by utilizing an attractive K3 surface $S_{[6 \hspace{1mm} 3 \hspace{1mm} 6]}$ as a sample, whose transcendental lattice has the intersection matrix as follows:
\begin{equation}
\begin{pmatrix}
6 & 3 \\
3 & 6 \\
\end{pmatrix}.
\end{equation}
\par The attractive K3 surface $S_{[6 \hspace{1mm} 3 \hspace{1mm} 6]}$ admits a genus-one fibration with a 3-section, but lacks a global section, with singularity type $E_6^3$, as discussed in \cite{Kimura2015} \footnote{As we will demonstrate in section \ref{sec4}, when the K3 surface $S_{[6 \hspace{1mm} 3 \hspace{1mm} 6]}$ has a genus-one fibration lacking a global section, the genus-one fibration must have a 3-section.}. The application of the Kneser--Nishiyama method to the attractive K3 surface $S_{[6 \hspace{1mm} 3 \hspace{1mm} 6]}$ shows that this K3 surface also admits an elliptic fibration with a section, as we now explain. $T_0$ for any attractive K3 surface admits a primitive embedding into the $E_8$ lattice by construction. Therefore, for any attractive K3, $T_0$ primitively embeds inside the Niemeier lattice $E_8^3$, where $T_0$ is embedded inside an $E_8$ lattice, and the orthogonal complement of this embedding includes $E_8^2$ \cite{BKW}. Therefore, an elliptic fibration with a section corresponding to this embedding contains two type $II^*$ fibers. For the attractive K3 surface $S_{[6 \hspace{1mm} 3 \hspace{1mm} 6]}$, the orthogonal complement of $T_0$ inside $E_8$ does not contain a root element; thus, the root lattice of the orthogonal complement of $T_0$ inside the lattice $E_8^3$ is $E_8^2$. Therefore, the attractive K3 surface $S_{[6 \hspace{1mm} 3 \hspace{1mm} 6]}$ admits an elliptic fibration with a section whose singularity type is $E_8^2$. The application of the Shioda--Tate formula \cite{Shiodamodular, Tate1, Tate2} applied to this fibration reveals that the Mordell--Weil rank of this fibration is two, and two U(1) factors are formed in F-theory on this fibration. 
\par Two F-theory limits, on the genus-one fibration without a section and the elliptic fibration that we just described, of M-theory on K3 surface $S_{[6 \hspace{1mm} 3 \hspace{1mm} 6]}$ yield very different theories; a discrete $\Z_3$ gauge symmetry and $\mathfrak{e}_6^3$ gauge algebra form \cite{Kimura2015} in the F-theory limit on the genus-one fibration lacking a global section, while $\mathfrak{e}_8^2\oplus \mathfrak{u}(1)^2$ gauge algebra forms in the F-theory limit on the elliptic fibration with a section, and a discrete gauge group does not form in the latter limit. 

\subsection{Heterotic perspective}
\label{subsec3.3}
As we discussed, on the F-theory side when the complex structure of a K3 surface is fixed, there is still a degree of freedom when choosing the fibration structures. On the 8D heterotic side on $T^2$, this freedom mainly corresponds to the choices of the vector bundles. The distinct F-theory limits of a single M-theory owing to the degree of freedom of fibration structures also yields different theories on the heterotic side. 
\par As limits of an identical M-theory on a K3 surface with a fixed complex structure, these different theories include both $E_8\times E_8$ and $SO(32)$ heterotic theories. This can be seen at the points in the F-theory moduli, where K3 surfaces are attractive. 
\par This follows from an observation that, given any attractive K3 surface, it admits two elliptic fibrations with a section, whose singularity types include $E_8^2$ and $D_{16}$ \cite{BKW}. $T_0$ computed in the context of the Kneser--Nishiyama method for an attractive K3 surface primitively embeds inside an $E_8$ of the Niemeier lattice $E_8^3$, and $T_0$ also primitively embeds inside $E_8$ of the Niemeier lattice whose root type is $E_8D_{16}$. The orthogonal complements of these embeddings of $T_0$ inside the Niemeier lattice $E_8^3$ and the Niemeier lattice whose root type is $E_8D_{16}$ yield two elliptic fibrations with a section for the attractive K3 surface whose singularity types include $E_8^2$ and $D_{16}$ \cite{BKW}.
\par When the orthogonal complement of $T_0$ inside $E_8$ does not contain a root element, the singularity types of the resulting elliptic fibrations are $E_8^2$ and $D_{16}$. When $T_0$ has a special structure, the orthogonal complement $T_0$ inside $E_8$ contains a root element. For such special situations, the singularity types of the two fibrations take one of the following three forms: $E_8^2A_2$ and $D_{16}A_2$, $E_8^2A_1^2$ and $D_{16}A_1^2$, and $E_8^2A_1$ and $D_{16}A_1$. The attractive K3 surface $S_{[2 \hspace{1mm} 1 \hspace{1mm} 2]}$ has two elliptic fibrations with a section whose singularity types are $E_8^2A_2$ and $D_{16}A_2$ \cite{Nish}. The attractive K3 surface $S_{[2 \hspace{1mm} 0 \hspace{1mm} 2]}$ has two elliptic fibrations with a section whose singularity types are $E_8^2A_1^2$ and $D_{16}A_1^2$. 
\par General attractive K3 surfaces with singularity type $E_8^2$ \footnote{The Shioda--Tate formula indicates that attractive K3 surfaces with singularity type $E_8^2$ have Mordell--Weil rank two. They correspond to special points in the complex structure moduli of K3 surfaces whose singularity type is $E_8^2$, at which the Mordell--Weil rank is enhanced to two.} that we just described above are obtained via the sum, $X_{[II^*, \hspace{1mm} 1, 1]}\cup X_{[II^*, \hspace{1mm} 1, 1]}$ of two copies of the extremal rational elliptic surfaces $X_{[II^*, \hspace{1mm} 1, 1]}$ glued together along the line of the argument, as in \cite{KRES, Kimura201810} \footnote{Some of the attractive K3 surfaces with the singularity type $E_8^2$ can be obtained via the sum, $X_{[II^*, \hspace{1mm} II]}\cup X_{[II^*, \hspace{1mm} II]}$ of two copies of the extremal rational elliptic surfaces $X_{[II^*, \hspace{1mm} II]}$.}. They stably degenerate into a pair of rational elliptic surfaces $X_{[II^*, \hspace{1mm} 1, 1]}$. 
\par The two elliptic fibrations with a section whose singularity types are $E_8^2A_2$ and $E_8^2A_1^2$ can be viewed as special limits of stable degenerations \cite{KRES}: As we stated, for the attractive K3 surface $S_{[2 \hspace{1mm} 1 \hspace{1mm} 2]}$ the singularity type becomes $E_8^2 A_2$, and this surface can be viewed as a special limit of the sum of two copies of the extremal rational elliptic surface $X_{[II^*, \hspace{1mm} II]}$ glued together, at which two type $II$ fibers collide and they are enhanced to a type $IV$ fiber \cite{KRES}. Here, $X_{[II^*, \hspace{1mm} 1, 1]}$ and $X_{[II^*, \hspace{1mm} II]}$ represent extremal rational elliptic surfaces with one type $II^*$ fiber and two type $I_1$ fibers, and an extremal rational elliptic surface with one type $II^*$ fiber and one type $II$ fiber, respectively, as mentioned in Section \ref{subsec2.1}. 
\par We also stated that the attractive K3 surface $S_{[2 \hspace{1mm} 0 \hspace{1mm} 2]}$ has an elliptic fibration with the singularity type $E_8^2A_1^2$ as discussed in \cite{Nish}. This K3 elliptic fibration can be viewed as a special limit of the reverse of the stable degeneration wherein two copies of extremal rational elliptic surfaces, $X_{[II^*, \hspace{1mm} 1, 1]}$, are glued together, at which two pairs of two type $I_1$ fibers collide, and they are enhanced to type $I_2$ fibers \cite{KRES}. 
\par It is clear that the two fibrations of an attractive K3 surface whose singularity types include $E_8^2$ and $D_{16}$ correspond to $E_8\times E_8$ and $SO(32)$ heterotic theories. They can be viewed as two different limits of a single M-theory on an attractive K3 surface. 
\par This can be viewed to reflect a well-known fact that the moduli spaces of $E_8\times E_8$ and $SO(32)$ heterotic strings are connected upon the toroidal compactification \cite{Narain1985, Narain1986}.

\subsection{Sample}
\label{subsec3.4}
To illustrate some of the discussions in sections \ref{subsec3.1} - \ref{subsec3.3}, we consider an example here. As mentioned in Section \ref{subsec2.1}, the complex structures of the attractive K3 surfaces are labeled by the triplets of integers, $2a, b, 2c$, where they satisfy the relation $a\ge c\ge b\ge 0$. This is owing to the fact that the complex structure of an attractive K3 surface is specified by the transcendental lattice \cite{PS-S1971, SI1977}, and the intersection matrix of the transcendental lattice can be expressed as a 2$\times$2 even symmetric integral matrix. 
\par We particularly consider an attractive K3 surface whose transcendental lattice has the following intersection matrix: $\begin{pmatrix}
12 & 0 \\
0 & 4 \\
\end{pmatrix}$. We denote this attractive K3 surface as $S_{[12 \hspace{1mm} 0 \hspace{1mm} 4]}$. Choosing this surface specifies a point in the 8D F-theory moduli on K3 surfaces; however, there still remains freedom when choosing the elliptic/genus-one fibration structures of the surface. 
\par We use the Kneser--Nishiyama method to determine the elliptic fibrations with a section of the surface $S_{[12 \hspace{1mm} 0 \hspace{1mm} 4]}$. $T_0$ of the K3 surface $S_{[12 \hspace{1mm} 0 \hspace{1mm} 4]}$ is a rank-six negative-definite lattice $A_3A_2(-4)$ \cite{BKW2013}. The orthogonal complements of this $T_0$ primitively embedded inside the Niemeier lattices classify the types of elliptic fibrations with a section of the surface $S_{[12 \hspace{1mm} 0 \hspace{1mm} 4]}$. We computed some of the fibrations with a section. 
\par One can embed $T_0=A_3A_2(-4)$ into Niemeier lattice $E_8^3$ as $A_3\subset E_8$, $A_2\subset E_8$, $(-4)\subset E_8$. The orthogonal complement of this embedding is $D_5E_6D_7$ which yields an elliptic fibration with a global section with singularity type $E_6D_7D_5$. This yields an extremal fibration. $\mathfrak{e}_6\oplus \mathfrak{so}(14)\oplus \mathfrak{so}(10)$ gauge algebra forms in the F-theory limit on this fibration. Because an extremal fibration has Mordell--Weil rank zero, the gauge group does not have a U(1) factor. 
\par One can also consider embedding $T_0=A_3A_2(-4)$ into a Niemeier lattice whose root type is $E_8D_{16}$ with $A_2\subset E_8$, $A_3\oplus (-4)\subset D_{16}$. The orthogonal complement of this yields an elliptic fibration with singularity type $E_6D_9A_3$. This is also an extremal fibration.
\par One can also consider embedding $T_0=A_3A_2(-4)$ into a Niemeier lattice whose root type is $E_7^2D_{10}$ with $A_2\subset E_7$, $A_3\oplus (-4)\subset D_{10}$. The orthogonal complement of this yields an elliptic fibration with singularity type $E_7A_5A_3^2$. This is also an extremal fibration. 
\par Another way of embedding is to embed $T_0=A_3A_2(-4)$ into a Niemeier lattice whose root type is $E_7^2D_{10}$ with $A_2\subset E_7$, $(-4)\subset E_7$, $A_3\subset D_{10}$. The orthogonal complement of this embedding yields an elliptic fibration with a section with singularity type $D_7D_5A_5A_1$. This yields another extremal fibration.  
\par One can also consider embedding $T_0=A_3A_2(-4)$ into a Niemeier lattice whose root type is $E_6D_7A_{11}$ with $A_2\subset E_6$, $A_3\oplus (-4)\subset D_7$. The orthogonal complement of this yields an elliptic fibration with singularity type $A_{11}A_3A_2^2$. This also results in an extremal fibration.
\par The authors in \cite{SZ} classified the extremal fibrations of the attractive K3 surfaces. The five extremal fibrations of the K3 surface $S_{[12 \hspace{1mm} 0 \hspace{1mm} 4]}$ that we computed using the Kneser--Nishiyama method are perfectly consistent with the results in Table 2 in \cite{SZ}. (They correspond to fibrations No.250, 253, 261, 167, 83 in Table 2 in \cite{SZ}, respectively.)
\par We can also deduce non-extremal fibrations. We consider embedding $T_0=A_3A_2(-4)$ into Niemeier lattice whose root type is $E_6D_7^2A_{11}$ with $A_3\oplus (-4)$ into $D_7$ and $A_2\subset A_{11}$. The orthogonal complement of this yields an elliptic fibration with a section whose singularity type is $E_6A_8A_3$. The Shioda--Tate formula \cite{Shiodamodular, Tate1, Tate2} applied to an elliptic K3 surface states that, the rank of the lattice generated by fiber components not meeting the zero-section and the rank of the Mordell--Weil group add to the Picard number of the K3 surface minus two. Because attractive K3 surface $S_{[12 \hspace{1mm} 0 \hspace{1mm} 4]}$ has Picard number twenty, and the singularity rank of the fibration that we just computed is seventeen, owing to the Shioda-Tate formula, we deduce the Mordell--Weil rank of the fibration with singularity type $E_6A_8A_3$ is one. F-theory on this elliptic fibration has one U(1) factor. $\mathfrak{e}_6\oplus \mathfrak{su}(9)\oplus \mathfrak{su}(4)\oplus \mathfrak{u}(1)$ gauge algebra forms in F-theory limit on this fibration.
\par Another non-extremal fibration is obtained via embedding $T_0$ into Niemeier lattice whose singularity type is $E_7^2D_{10}$, with $A_3\subset E_7$, $(-4)\subset E_7$, $A_2\subset D_{10}$. The orthogonal complement of this yields an elliptic fibration with a section whose singularity type is $D_7D_5A_3A_1^2$. Using an argument similar to that we discussed previously, this fibration also has Mordell--Weil rank one, and U(1) is formed in F-theory on this fibration. 
\par The seven elliptic fibrations of K3 surface $S_{[12 \hspace{1mm} 0 \hspace{1mm} 4]}$ correspond to seven different F-theory limits of an identical M-theory on $S_{[12 \hspace{1mm} 0 \hspace{1mm} 4]}$, and different non-Abelian gauge groups form, and the numbers of U(1) factors formed in the theories are different in these limits.

\par Using an argument similar to that given in \cite{KRES, Kimura201810}, one finds that an elliptic fibration with singularity type $E_7A_5A_3^2$ of the K3 surface $S_{[12 \hspace{1mm} 0 \hspace{1mm} 4]}$ is obtained as a deformation of the sum, $X_{[III^*, \hspace{1mm} 2, 1]}\cup X_{[6, \, 3, \, 2, \, 1]}$, of two extremal rational elliptic surfaces, $X_{[III^*, \hspace{1mm} 2, 1]}$ and $X_{[6, \, 3, \, 2, \, 1]}$, where two $I_2$ fibers collide, and they are enhanced to a type $I_4$ fiber, and a type $I_3$ fiber and a type $I_1$ collide and they are enhanced to a type $I_4$ fiber. K3 surface $S_{[12 \hspace{1mm} 0 \hspace{1mm} 4]}$ with an elliptic fibration $f:S_{[12 \hspace{1mm} 0 \hspace{1mm} 4]}\rightarrow \P^1$ with singularity type $E_7A_5A_3^2$ stably degenerates into two extremal rational elliptic surfaces $X_{[III^*, \hspace{1mm} 2, 1]}$ and $X_{[6, \, 3, \, 2, \, 1]}$, at which heterotic/F-theory duality becomes precise.

\section{Lattice theoretic application to discrete gauge groups}
\label{sec4}
A discrete $\Z_n$ gauge group arises in F-theory on a genus-one fibration with an $n$-section, but lacks a global section \cite{MTsection}. Here, we discuss the application of the lattice theory approach to a discrete gauge symmetry in F-theory.
\par Analyzing lattice structures of the geometry imposes a constraint on the degree of a multisection possessed by a genus-one fibration. In the context of F-theory, this analysis constrains the degree of the discrete gauge symmetry. 
\par Here, we utilize a lemma in \cite{Keum}. The lemma in \cite{Keum} is a statement on a relation that holds between the determinant of the Picard lattice of a genus-one fibered K3 surface $S$ and the determinant of the Picard lattice of the Jacobian fibration $J(S)$. On this note, we mainly focus on the application of the lemma in \cite{Keum} to attractive K3 surfaces. With some appropriate modifications to our argument, a similar discussion applies to K3 surfaces that have other Picard numbers. 
\par When applied to the attractive K3 surfaces, the lemma in \cite{Keum} that we utilize here states \footnote{The determinants of the Picard lattice and the transcendental lattice of a K3 surface have the same absolute values \cite{Nik}.} that: given a genus-one fibered K3 surface $S$ of degree $n$, i.e., the genus-one fibration has an $n$-section, and $J(S)$ is its Jacobian fibration, then the following relation holds between the determinants of the transcendental lattices of K3 surface $S$ and the Jacobian fibration $J(S)$:
\begin{equation}
\label{Picard formula in 4}
{\rm det}\, T_S=n^2 {\rm det} \, T_{J(S)}
\end{equation}
\par This relation, together with the existence of the Jacobian fibration of a K3 genus-one fibration, imposes on the degree $n$ of a multisection of a genus-one fibration. We use the attractive K3 surface $S_{[6 \hspace{1mm} 3 \hspace{1mm} 6]}$ as a sample, and we show how the approach applies to F-theory and a discrete gauge symmetry on this attractive K3. 
\par The transcendental lattice of attractive K3 surface $S_{[6 \hspace{1mm} 3 \hspace{1mm} 6]}$ has the intersection matrix $\begin{pmatrix}
6 & 3 \\
3 & 6 \\
\end{pmatrix}$; therefore, it has the determinant 27. Owing to the equality (\ref{Picard formula in 4}), if attractive K3 $S_{[6 \hspace{1mm} 3 \hspace{1mm} 6]}$ admits a genus-one fibration, the square of the degree of a multisection that the fibration possesses must divide 27. Thus, a unique possibility of the degree of a multisection is 3; if $S_{[6 \hspace{1mm} 3 \hspace{1mm} 6]}$ admits a genus-one fibration lacking a global section, then it must have a 3-section. 
\par Every genus-one fibered K3 surface has the Jacobian fibration. Using this fact, we can impose further constraints on the genus-one fibrations. Because the degree of a genus-one fibration of $S_{[6 \hspace{1mm} 3 \hspace{1mm} 6]}$ must be three if it admits a genus-one fibration without a section, the determinant of the transcendental lattice --or more concisely the {\it discriminant}, of the Jacobian fibration is then 3 owing to equation (\ref{Picard formula in 4}). It is known that 3 is the smallest value among the discriminants of the attractive K3 surfaces, and an attractive K3 surface with discriminant 3 is unique and isomorphic to the attractive K3 surface $S_{[2 \hspace{1mm} 1 \hspace{1mm} 2]}$. An explanation of this in the context of string theory can be found in \cite{KimuraMizoguchi}. From these results, we conclude that the Jacobian of any genus-one fibration of $S_{[6 \hspace{1mm} 3 \hspace{1mm} 6]}$ lacking a global section must be the attractive K3 surface $S_{[2 \hspace{1mm} 1 \hspace{1mm} 2]}$. 
\par Because the singularity type of a genus-one fibration and that of the Jacobian fibration are identical, this means that a genus-one fibration of $S_{[6 \hspace{1mm} 3 \hspace{1mm} 6]}$ has a singularity type identical to an elliptic fibration with a section of $S_{[2 \hspace{1mm} 1 \hspace{1mm} 2]}$. The author in \cite{Nish} classified the types of elliptic fibrations of the K3 surface $S_{[2 \hspace{1mm} 1 \hspace{1mm} 2]}$, and determined that the K3 surface $S_{[2 \hspace{1mm} 1 \hspace{1mm} 2]}$ has six fibration types with singularity types: $E_6^3$, $E_8^2 A_2$, $D_{16}A_2$, $D_7A_{11}$, $E_7D_{10}$, and $A_{17}$. 
\par We deduce from these results that if $S_{[6 \hspace{1mm} 3 \hspace{1mm} 6]}$ admits a genus-one fibration without a section, then it must have a 3-section, and the singularity type of the genus-one fibration must be one of the six types: $E_6^3$, $E_8^2 A_2$, $D_{16}A_2$, $D_7A_{11}$, $E_7D_{10}$, and $A_{17}$. 
\par In the string theoretic language, this implies that if a discrete gauge group arises in F-theory on $S_{[6 \hspace{1mm} 3 \hspace{1mm} 6]}$, then it must be a $\Z_3$ symmetry, and the continuous gauge group factor formed in this theory must be one of the following: $E_6^3/\Z_3$, $E_8^2\times SU(3)$, $SO(32)\times SU(3)/\Z_2$, $SO(14)\times SU(12)/\Z_4$, $E_7\times SO(20)/\Z_2\times U(1)$, $SU(18)/\Z_3\times U(1)$ \footnote{The Mordell--Weil groups of the elliptic fibration types of the K3 surface of discriminant 3, $S_{[2 \hspace{1mm} 1 \hspace{1mm} 2]}$ were obtained in \cite{Nish}. These are presented in Table \ref{tab fibrations of surface discriminant 3 in 2.2}. The Mordell--Weil rank yields the number of U(1) factors formed in F-theory on Jacobian fibration \cite{MV2}. The fundamental group $\pi_1(G)$ of the gauge symmetry, $G$, formed in F-theory on an elliptic fibration, is isomorphic to the torsion part of the Mordell--Weil group of the elliptic fibration, as discussed in \cite{AspinwallGross, AMrational, MMTW}.}.
\par It was shown in Section 4 in \cite{Kimura2015} that attractive K3 surface $S_{[6 \hspace{1mm} 3 \hspace{1mm} 6]}$ admits a genus-one fibration with a 3-section, but lacks a global section, with the singularity type $E_6^3$. Therefore, we confirm that our analysis is consistent with the result obtained in \cite{Kimura2015}. In this study, we do not determine whether the attractive K3 surface $S_{[6 \hspace{1mm} 3 \hspace{1mm} 6]}$ admits another genus-one fibration without a section. 
\par More generally, a similar argument applies to an attractive K3 surface with discriminant $n^2\, m$, where $n$ is a prime and $m$ is square-free. If this K3 surface admits a genus-one fibration lacking a global section, it must have an $n$-section and the Jacobian surface must have the discriminant $m$. If the complex structure of the Jacobian surface can be determined, then the Kneser--Nishiyama method applies to the Jacobian surface and one can constrain the possible singularity types to finite candidates. Thus, one learns that if a discrete gauge symmetry arises in F-theory on an attractive K3 surface with discriminant $n^2\, m$, it must be a discrete $\Z_n$ symmetry and one can limit candidates for continuous gauge group formed in the theory to finite possibilities. 

\par We considered attractive K3 surface $S_{[12 \hspace{1mm} 0 \hspace{1mm} 4]}$ in Section \ref{subsec3.4}. We would like to apply the previously discussed topic to this attractive K3 surface. The attractive K3 surface $S_{[12 \hspace{1mm} 0 \hspace{1mm} 4]}$ has discriminant 48; therefore, if it admits a genus-one fibration that lacks a global section, then the fibration has either a 4-section or a bisection. 
\par If a genus-one fibration without a section of the surface has a 4-section, then owing to the equation (\ref{Picard formula in 4}), the Jacobian surface must have discriminant 3. Thus, the Jacobian K3 surface is isomorphic to $S_{[2 \hspace{1mm} 1 \hspace{1mm} 2]}$. Similar to what we have previously argued, the singularity type of the genus-one fibration with a 4-section must be one of the six types: $E_6^3$, $E_8^2 A_2$, $D_{16}A_2$, $D_7A_{11}$, $E_7D_{10}$, and $A_{17}$. 
\par If a genus-one fibration without a section of the surface has a bisection, from (\ref{Picard formula in 4}), we deduce that the Jacobian surface of this fibration must have the discriminant 12. 
\par We do not determine whether or not attractive K3 surface $S_{[12 \hspace{1mm} 0 \hspace{1mm} 4]}$ actually admits a genus-one fibration without a section with a bisection or with a 4-section. 

\section{Conclusions}
\label{sec5}
In the F-theory moduli, compactification spaces admit several elliptic/genus-one fibrations, and we discussed how this can be viewed from M-theoretic and heterotic perspectives when the compactification spaces are K3 surfaces in the F-theory moduli. The different fibration structures of an elliptic K3 surface correspond to the different small-fiber F-theory limits of a single M-theory. 
\par Elliptic K3 surfaces generally admit both elliptic fibrations with a section and a genus-one fibration without a section. This physically means that considering F-theory on an elliptic fibration with a global section and on a genus-one fibration lacking a global section, starting from an M-theory on an elliptic K3 surface, one obtains both an F-theory limit wherein a discrete gauge symmetry does not form, and another F-theory limit wherein a discrete gauge symmetry forms. 
\par We analyzed the elliptic fibrations with a section of elliptic K3 surfaces, particularly when the surfaces are attractive, applying a mathematical method known as the Kneser--Nishiyama method. This method studies the lattice structures of K3 surfaces and extracts information of the gauge groups formed in F-theory on K3 surfaces. We studied the points in the F-theory moduli of K3 surfaces where the surfaces are attractive, and we discussed the M-theoretic and heterotic perspectives of these points. 
\par Different fibration structures of an attractive K3 surface include, on the heterotic side, both $E_8\times E_8$ and $SO(32)$ heterotic theories. This implies that these theories originate from an identical M-theory, and this can be viewed as a reflection of the fact that the moduli spaces of $E_8\times E_8$ and $SO(32)$ heterotic strings are connected upon toroidal compactification. 
\par We also discussed some connections to the stable degeneration limit for attractive K3 surfaces.
\par The Kneser--Nishiyama method classifies the elliptic fibrations with a section of complex K3 surfaces with Picard numbers greater than or equal to 12 \cite{SchShio}. However, this method does not classify genus-one fibrations without a section of such surfaces. It would be interesting to search for a method to classify genus-one fibrations that lack a global section.

\par Analyzing the lattice structures of the attractive K3 surfaces, we explained that the discriminants and the existence of the Jacobian fibrations limit the possible degrees of a discrete gauge group and candidates for the continuous gauge group formed in F-theory on an attractive K3 surface. For some attractive K3 surfaces that have specific complex structures, these conditions uniquely determine the degree of a discrete gauge group if a discrete gauge group is formed in F-theory on the attractive K3 surface.
\par Elliptic fibrations of the attractive K3 surface $S_{[12 \hspace{1mm} 0 \hspace{1mm} 4]}$ include an elliptic fibration with a section with singularity type $E_6D_9A_3$, as discussed in Section \ref{subsec3.4}. By employing reasoning that is similar to that given in \cite{Kimura201810, Kimura201902}, the gauge group formed in the heterotic dual of the 8D F-theory model on this fibration does not allow for a perturbative interpretation. Because every attractive K3 surface admits an elliptic fibration with a section whose singularity type includes $E_8^2$ \cite{BKW}, $S_{[12 \hspace{1mm} 0 \hspace{1mm} 4]}$ admits an elliptic fibration with singularity type $E_8^2$. The equation for the fibration with singularity type $E_8^2$ of $S_{[12 \hspace{1mm} 0 \hspace{1mm} 4]}$ presumably admits a transformation to an elliptic fibration with singularity type $E_6D_9A_3$. Using the argument given in \cite{Kimura201810, Kimura201902}, from the perspective of non-geometric heterotic strings \cite{MMS, MM, GJ1412, LMV1508, FGLMM1603, MS1609, GLMM1611, FM1708, FGLMM1712, Clingher2018, Kimura201810, Kimura201902, Clingherduality}, the gauge groups that do not allow for perturbative interpretation are understood as the nonperturbative effect of the insertion of 5-branes on the heterotic side. Through a transformation between two elliptic fibrations, we expect that the gauge group formed in the heterotic dual of F-theory on the fibration with singularity type $E_6D_9A_3$ can be understood to result from the nonperturbative effect of the insertion of 5-branes.

\section*{Acknowledgments}

We would like to thank Shun'ya Mizoguchi and Shigeru Mukai for discussions.

\end{document}